# Mutual Coupling in Compact Orthomode Transducers

Matthew A. Morgan – matt.morgan@nrao.edu

National Radio Astronomy Observatory
1180 Boxwood Estate Rd.
Charlottesville, VA. 22903

**Abstract**

The scattering parameters of generalized compact orthomode transducers using azimuthally-distributed field probes in a dual-mode waveguide are analyzed. Theoretical expressions constraining the mutual coupling between the probes are derived and evaluated for three- and four-probe orthomode transducers with and without a coaxial reference port for calibration injection. The mutual coupling is shown to be identically zero or cancel coherently in all cases, suggesting that radiometric receivers with the best possible system noise temperature may be realized using these topologies.

## I. Introduction

Recent experiments have shown that compact orthomode transducers with state-of-the-art polarization performance may be constructed using an arbitrary number of field probes in a backshorted dual-mode waveguide, combined with stable receiver design and numerical calibration [1]. We now extend that work by analyzing the scattering parameters of the generalized topology for an arbitrary number of output channels, $N$. Special attention will be given to the mutual coupling between the field probes, as back-emitted noise from the low-noise amplifiers coupling into adjacent channels may tend to increase the noise temperature in otherwise high-performance receivers [2].

## II. Analysis

The general topologies with which we will be concerned are shown in Fig. 1, wherein the probes extending into a hollow waveguide on the left and into an overmoded coaxial waveguide on the right. The structure is assumed to be symmetric about the probe axes, but the shape of the probes and of the waveguide's cross-section is otherwise arbitrary. In fact, the analysis applies equally well to turnstile junctions which have no probes at all. The figure shows $N=3$ probes (or arms, in a turnstile junction), which will be an important special case, but for now we consider $N$ also as being arbitrary. Even with these limited assumptions, the scattering parameters of this network may be derived in part by application of the rules of symmetry and reciprocity.

First, we assert that the response of any probe to a linear polarization which is perpendicular to its axis must be identically zero. This is easily seen because such a field pattern permits a symmetry plane corresponding to a virtual electric-wall boundary condition to be drawn through the structure and through the probe's axis, effectively shorting out that port. Therefore, the probe only responds to linear polarization components parallel to its axis. Mathematically, the net response of probe $P$ to any arbitrary linear polarization $L$ is given by

$$s_{LP} = s_{PL} = Ae^{j\phi}\cos\theta \qquad (1)$$

where $A$ is the scalar amplitude, $\phi$ is the relative phase, and $\theta$ is the angle of the peak input E-field relative to the axis of the probe. We assume the orthomode transducer has been impedance-matched, or that $s_{LL}=0$. Furthermore, the symmetry of the structure is also sufficient to ensure that no cross-coupling between the linear polarizations can occur [1], resulting in the stronger statement

$$s_{L_iL_k} = 0 \qquad (2)$$

for all $i$ and $k$.

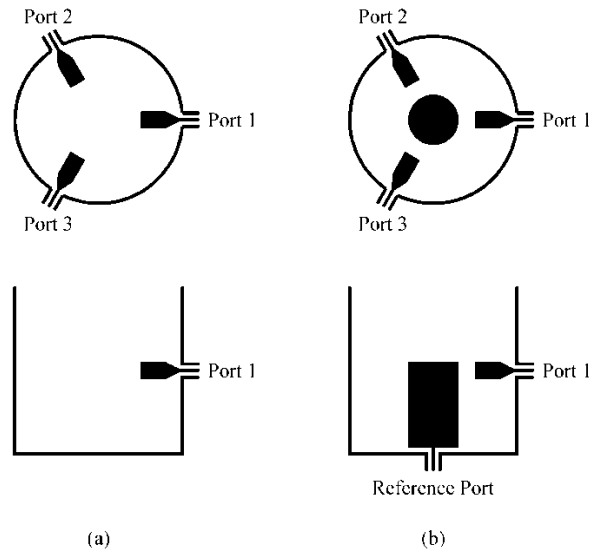

Fig. 1. Cross-section of compact orthomode transducers comprising azimuthally-distributed field probes in a dual-mode waveguide. a) Original configuration. b) Modified configuration with a coaxial reference port for injection of calibration signals. $N=3$ probes are shown above, but for much of the analysis in this paper, the number of probes is arbitrary.

Since the structure is lossless, the scattering parameter matrix is unitary. For the terms in column $L_1$, then, we may write

$$\left|s_{L_1 L_1}\right|^2 + \left|s_{L_2 L_1}\right|^2 + \sum_{k=1}^{N}\left|s_{P_k L_1}\right|^2 = 1 \quad (3a)$$

$$\sum_{k=1}^{N} A^2 \cos^2\left(\theta + \tfrac{2\pi(k-1)}{N}\right) = \tfrac{1}{2} A^2 N = 1 \quad (3b)$$

$$\therefore A = \sqrt{\tfrac{2}{N}}. \quad (3c)$$

In effect, this shows that the net power delivered by a single probe into the two linear polarizations is $2/N$.

Additionally, the inner product of row $P_1$ and column $L_1$ of the unitary scattering parameter matrix may be written as

$$s_{P_1 L_1} s_{L_1 L_1}^* + s_{P_1 L_2} s_{L_2 L_1}^* + \sum_{k=1}^{N} s_{P_1 P_k} s_{P_k L_1}^* = 0 \quad (4a)$$

$$\left[\Gamma \cos\theta + \sum_{k=2}^{N} C_k \cos\left(\theta + \tfrac{2\pi(k-1)}{N}\right)\right]\sqrt{\tfrac{2}{N}} e^{-j\phi} = 0 \quad (4b)$$

$$\Gamma \cos\theta + \sum_{k=2}^{N} C_k \cos\left(\theta + \tfrac{2\pi(k-1)}{N}\right) = 0 \quad (4c)$$

where $\Gamma$ is the individual probe's reflection coefficient and $C_k$ is the mutual coupling between probe 1 and probe $k$ (or, more generally, between any two probes $k$-1 positions apart). This is required to hold for all consistent choices of $L_1$ and $L_2$, and hence for all $\theta$. Additionally, symmetry requires that $C_k=C_{N+2-k}$ for all $k$, and we may simplify (4c) to

$$\Gamma + \sum_{k=2}^{N} C_k \cos\left(\tfrac{2\pi(k-1)}{N}\right) = 0. \quad (5)$$

Thus far, we have not specified whether the orthomode transducer includes a coaxial reference port or not, as depicted comparatively in Fig. 1. Since the TEM coaxial mode is mathematically orthogonal to the linear polarizations, the terms for the coaxial port, if present, would be identically zero, and the above expressions would not change. In what follows, however, we must take this distinction into account.

If there is not a coaxial port, then the square magnitude of column $P_1$ of the scattering matrix may be written as

$$\left|s_{L_1 P_1}\right|^2 + \left|s_{L_2 P_1}\right|^2 + \sum_{k=1}^{N}\left|s_{P_k P_1}\right|^2 = 1 \quad (6a)$$

$$\tfrac{2}{N}\cos^2\theta + \tfrac{2}{N}\sin^2\theta + \left|\Gamma\right|^2 + \sum_{k=2}^{N}\left|C_k\right|^2 = 1 \quad (6b)$$

$$\left|\Gamma\right|^2 + \sum_{k=2}^{N}\left|C_k\right|^2 = 1 - \tfrac{2}{N}. \quad (6c)$$

On the other hand, if the structure does contain a coaxial reference port, located concentrically between the probes and exciting them via a TEM mode in the waveguide, the situation is somewhat changed. By symmetry, this additional port remains isolated from the two linear polarization inputs ($s_{LX}=s_{XL}=0$, where the $X$ subscript denotes the coaxial port). Also, like the linear modes, we assume the TEM mode is impedance-matched ($s_{XX}=0$). Thus, energy from the coaxial port couples into all $N$ probes equally, and

$$\left|s_{PX}\right| = \left|s_{XP}\right| = \sqrt{\tfrac{1}{N}}. \quad (7)$$

This contributes an additional term of $1/N$ to the above summation, allowing us to generalize (6c) as

$$\left|\Gamma\right|^2 + \sum_{k=2}^{N}\left|C_k\right|^2 = \begin{cases} 1-\tfrac{2}{N} & \text{without TEM probe} \\ 1-\tfrac{3}{N} & \text{with TEM probe} \end{cases}. \quad (8)$$

Together, (5) and (8) provide powerful constraints on the possible values of the scattering parameters of an input-matched orthomode transducer, regardless of the details of its design.

### III. Special Cases

We now use (5) and (8) to evaluate the mutual coupling in two special cases. In the most conventional case, where $N=4$, the result of (5) is that

$$\Gamma = C_3, \quad (9)$$

or, in other words, that the reflection coefficient looking into the waveguide from any one probe is equal – in magnitude and in phase – to the coupling coefficient with the opposing probe. Since these probes are nominally differenced to reconstruct the original polarized signal, either with a passive combiner or numerical processing, the mutually coupled signals cancel out. Further, since we have $C_2=C_4$, the coupling into the orthogonal channels also cancel. Thus, for $N=4$, the mere fact that mutual coupling may be present does not directly impact the potential receiver noise temperature.

The other special case, where $N=3$, is in some ways even more interesting. Recognizing that $C_2=C_3=C$, we substitute into (5) and (8) to find

$$\left|\Gamma\right|^2 = \left|C\right|^2 = \begin{cases} \tfrac{1}{9} & \text{without TEM probe} \\ 0 & \text{with TEM probe} \end{cases}. \quad (10)$$

Thus, if there is no coaxial TEM port, the mutual coupling and individual reflection coefficients are each 1/9, and it can be shown, as it was in the four-probe case, that all coupled noise signals must cancel.

Somewhat surprising is the result for $N=3$ in which there is a coaxial reference port. In this instance, the probes are both isolated ($C=0$) and individually matched ($\Gamma=0$). The energy from any signal impressed upon a single probe is divided among the waveguide ports ($2/3^{rds}$ to the linear outputs, $1/3^{rd}$ to the coaxial), but no energy is coupled into either of the other probes. The reference port presents a common-mode termination to the probes which may be considered as playing a role similar to the balancing resistor in a Wilkinson Power Divider [3], providing intrinsic isolation between the outputs of the device without incurring any additional loss in the desired signal path.

## IV. Discussion and Conclusions

This analysis has shown that radiometric receivers may be constructed using either orthomode transducer topology in Fig. 1, with either three or four probes, without suffering any direct penalty due to noise scattering. The three-probe topology with calibration port is doubly-protected, in that its mutual coupling is nominally zero, and what coupling is present due to manufacturing tolerances should be largely coherent in all channels and cancel.

However, the topology in Fig. 1b with $N=3$ does have other subtle advantages that are not directly apparent from this analysis. With the addition of a concentric TEM port in the waveguide, we have the opportunity to introduce signals into the system which pass through the same active electronics as the linearly polarized signals from the sky, but in a mode which is inherently decoupled from them. That is, since the TEM and two $TE_{11}$ modes are mutually orthogonal, the numerical reconstruction of any one of them in the backend processor intrinsically excludes the other two.

This extra port might easily be used for system calibration by noise injection, or for the insertion of a local oscillator into SIS-mixer front-ends. In so doing, we eliminate the need for a coupler or optical beam splitter to serve this purpose, and consequently avoid the insertion loss of these components prior to amplification. Neither a two-arm orthomode transducer (such as a quad-ridge) or a four-arm design with passive combiners (such as the many turnstile-junction based approaches) can do this. In those instances the number of *active* receiver channels is still only two, which have insufficient degrees of freedom to carry three numerically orthogonal signals, namely the two linear polarizations and the reference signal. Four-probe topologies without passive combiners meet the necessary criteria, but at the cost of additional electronic hardware that has no obvious advantages (aside, perhaps, from the greater intuitiveness of their operation).

But the implications of this result go deeper still. For it illustrates that the minimal coupling between adjacent field probes need not be bound by proximity. If a similarly straightforward technique could be found to nullify the mutual coupling between antennae in a close-packed two-dimensional array, then the full potential of radiometric Phased-Array Feeds with uncompromised noise performance might finally be realized.

## Acknowledgment

The author thanks his colleagues, Rick Fisher for helping to develop the compact digital orthomode transducer, and Marian Pospieszalski for emphasizing the importance of mutual coupling in low-noise multi-input receivers.